\documentclass[aps,pra,twocolumn,floatfix,superscriptaddress]{revtex4}  
\usepackage{float}
\usepackage{graphicx}
\usepackage{amssymb}
\usepackage{dsfont}
\usepackage{amsmath}  
\usepackage{epstopdf}
\usepackage{color}

\begin{document}

\def\ket#1{|#1\rangle} 
\def\bra#1{\langle#1|}
\def\av#1{\langle#1\rangle}
\def\dkp#1{\kappa+i(\Delta+#1)}
\def\dkm#1{\kappa-i(\Delta+#1)}
\def\pp{{\prime\prime}}
\def\ppp{{\prime\prime\prime}}
\def\w{\omega}
\def\k{\kappa}
\def\D{\Delta}
\def\wp{\omega^\prime}
\def\wpp{\omega^{\prime\prime}}

\def\u{\color{blue}}
\def\r{\color{red}}
\def\b{\color{black}}

\title{One- and Two-Photon Scattering by Two Atoms in a Waveguide}
\author{William Konyk and Julio Gea-Banacloche}
\affiliation{Department of Physics, University of Arkansas, Fayetteville, AR 72701}

\date{\today}

\begin{abstract}
%No abstract yet!
We consider the interaction of one- and two-photon pulses in a waveguide with two two-level systems (TLS) that are also able to interact directly either through an exchange- or a dipole-type interaction.  We focus on the system's transport properties and show how the presence of a second TLS increases the control options, especially when direct interactions are also allowed.  We also obtain, within a Markov (long pulse) approximation, exact results for the nonlinear or entangled terms that arise in the two-photon case, and discuss both their potential applications and ways to minimize their effects. 
\end{abstract}
\maketitle

\section{Introduction and summary}
The study of the interaction of few-photon pulses with atoms coupled to a waveguide has recently attracted a great deal of interest \cite{shen1,gauthier,hach,fan1,mork1,combes,gritsev,garciavidal1,zheng2,zheng3,sanders,garciavidal2,laakso,epj,mork2,fan15,zheng4,zubairy1,cirac,us,roulet,mirza1,mirza2,interactions}, motivated in part by the experimental progress made in related systems for quantum computing (such as superconducting qubits).  From a theoretical point of view, the unidimensional geometry makes numerical calculations relatively straightforward, and a number of analytical results for various configurations do exist as well (see, particularly, for multiple atoms and photons \cite{gritsev,fan15,laakso,zheng4,cirac}).  

In a recent paper \cite{us}, we have studied the scattering of multimode Fock states by a single atom, in a one-dimensional configuration, in the absence of external losses. We derived a result that allows one to calculate the scattered field state explicitly, in closed form, as a series of $N$ (for initially uncorrelated states) or, at most, $2N$ nested integrals, where $N$ is the incident number of photons.  (Interestingly, an equivalent form of this result was published almost simultaneously by Roulet and Scarani \cite{roulet}, working in the Heisenberg picture.)  Our goal here is to extend this approach to the two-atom problem, where the presence of the second atom may be expected to offer additional control over the final scattering state. Unlike most previous treatments (but see \cite{interactions} for a very recent exception), we allow, in principle, for a direct interaction between the atoms (or more generally, two-level systems), of the kind that one might observe, for example, in Rydberg-atom or quantum dot systems. 

Our focus is on the system's transport properties and the nonlinear effects arising from the two-photon interaction, including the generation of two-photon entangled states and conditional phase shifts.  As we did before, we try to provide explicit expressions for the field (and the atoms') states wherever possible.  We find, however,  that manageable analytic expressions of some generality are really only possible for the two-photon case if a ``Markovian'' approximation is made (in essence, assuming that the pulse's length is large compared to the separation between the atoms).  

Our paper is organized as follows.  We start, in Section II, with the single-photon, two-atom problem, where the Markovian approximation is, in fact, not necessary to solve for the final scattering state (as already shown in, e.g., \cite{garciavidal1,zheng2,zubairy1}).  We find this to be true also when direct interactions are allowed, at least for the models we consider here.  We study in detail the system's transmission properties and show how these are modified, and additional control over the final state is possible, when a dipole-dipole or exchange interaction term is available. Our results in this section are presented in the frequency domain.

In Section III we invoke the Markov approximation and solve for the evolution of the atom and field states by an extension of the approach we used in \cite{us}.  Here most of the results for the final two-photon ``wavefunctions'' are in the time domain. We present explicit expressions for the nonlinear, entangled terms that arise from different configurations (single standing wave, or two traveling photons arriving from the same or opposite directions). We discuss their size, the way the affect the transmission properties of the system, and possible applications, including photon sorting and conditional phase gates.  Section IV contains a brief summary of our results and our conclusions.

\section{One photon, two atoms}

\subsection{Setup and notation}

The situation we will consider here is schematically depicted in Figure 1.  The two atoms are located at $z_1 = -a/2$ and $z_2 = a/2$.  The left-traveling modes are represented by the canonical creation and annihilation operators $\hat a^\dagger_\omega$ and $\hat a_\omega$, satisfying $[\hat a_\omega,a^\dagger_{\omega^\prime}] = \delta(\omega-\omega^\prime)$, and likewise the right-traveling modes have operators $\hat b^\dagger_\omega$ and $\hat b_\omega$. Throughout the paper, the symbol $\omega$ will denote not the actual field frequency, but the difference between this and the ``carrier'' frequency $\omega_F$ (assumed to be the same for both sets of modes).  Thus, the positive-frequency part of the right- and left-traveling field operators are
\begin{align}
E_a^{(+)}(t,z) &= \sqrt{\frac{\hbar \omega_F}{2\epsilon_0}}\int e^{-i(\omega_F + \omega)t + i(k_F +\omega/c)z} \hat a_\omega \, d\omega \cr
E_b^{(+)}(t,z) &= \sqrt{\frac{\hbar \omega_F}{2\epsilon_0}}\int e^{-i(\omega_F + \omega)t - i(k_F +\omega/c)z} \hat b_\omega \, d\omega
\label{e1}
\end{align}  
where $k_F = \omega_F/c$.  The symbol $c$ is the phase velocity in the waveguide, which we will take to be equal to the group velocity over the range of frequencies considered (no dispersion).  We will also take the pulses to be sufficiently narrow-band for the coupling constants to be approximately independent of $\omega$ and depend only on $\omega_F$.

\begin{figure}
\includegraphics[width=8cm]{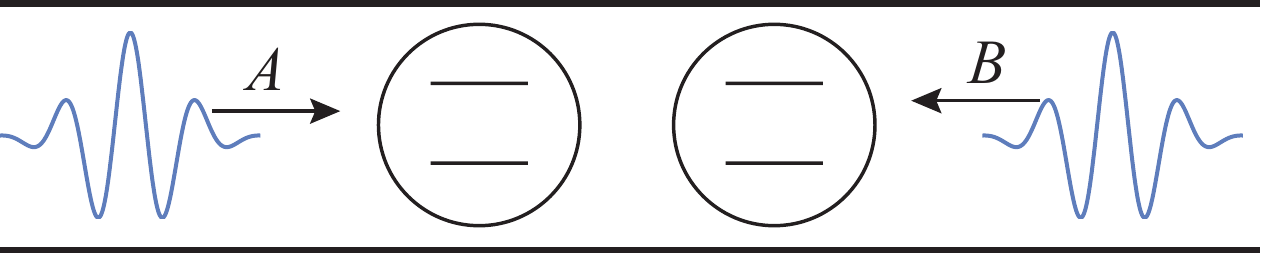}
\caption{\label{fig1} The two atoms in the waveguide interacting with oppositely-directed running-wave modes $A$ and $B$.}
\end{figure}

In terms of $\omega$, we can write the two-photon state as
\begin{align}
\ket{\psi} &= \frac{1}{\sqrt{2}} \int \int d\omega_1\, d\omega_2 \tilde f(\omega_1,\omega_2) \hat a_{\omega_1}^\dagger \hat a_{\omega_2}^\dagger \ket 0 \cr
&= \frac{1}{\sqrt{2}} \int \int dt_1\, dt_2  f(t_1,t_2) \hat\phi_a^\dagger (t_1) \hat\phi_a^\dagger (t_2) \ket 0
\end{align}
where $\tilde f(\omega_1,\omega_2)$ (square-normalized to 1) is the shifted spectrum, centered around $0$ instead of $\omega_F$;  its two-dimensional Fourier transform is the two-photon ``wavefunction'' $f(t_1,t_2)$ (basically, the pulse's slowly-varying envelope); and the operator 
\begin{equation}
\hat\phi_{a}^\dagger(t)= \frac{1}{\sqrt{2\pi}} \int d\omega e^{i\omega t} \hat a^\dagger_\omega,
\label{ne3}
\end{equation}
which satisfies the commutation relations $[\hat\phi_a(t),\hat\phi_{a}^\dagger(t^\prime)]= \delta(t-t^\prime)$ may be said to create a photon at the instant $t$.

The two atoms are assumed to be identical in their coupling to the field, denoted below by the constant $g$, and their resonant frequency $\omega_A$.  Defining the detuning $\delta = \omega_F-\omega_A$, the atom-field interaction Hamiltonian can be written as

\begin{align}
H_I = &\hbar g\sum_{i=1,2} \big[e^{ik_F z_i}\hat\phi_a(t-z_i/c) + e^{-ik_F z_i}\hat\phi_b(t+z_i/c) \big] \cr
&\times e^{-i\delta t}\ket{e}_i\bra{g} + H.c.
\label{e3}
\end{align}
in a suitable interaction picture.

In the same picture, we may include a direct atom-atom interaction Hamiltonian of the form
\begin{equation}
H_A = \hbar\Delta \bigl(\ket{eg}\bra{ge} + \ket{ge}\bra{eg}\bigr) + \hbar\beta\ket{ee}\bra{ee}
\label{e4}
\end{equation}
The term proportional to $\Delta$ is an exchange- (or F\"orster-) type interaction \cite{forster}, and the term proportional to $\beta$ represents a detuning of the doubly excited state due to the dipole interaction.  Both terms are often used to model interactions between neutral atoms \cite{deutsch, zubairy3, Draayer}, but could appear also in other contexts, such as when the two-level systems are closely-spaced quantum dots.  Note that, in a very recent paper \cite{interactions}, Liao, Nha and Zubairy have considered a model for the dipole-dipole interaction mediated by both the waveguide and non-waveguide modes, which does not seem to be reducible to the form (\ref{e4}), but nonetheless shares some of the features we will discuss below. 

The form of the Hamiltonian (\ref{e3}) immediately suggests that it would be advantageous to introduce the linear combinations $\hat c_\omega = (\hat a_\omega+\hat b_\omega)/\sqrt 2$ and $\hat d_\omega = (\hat a_\omega-\hat b_\omega)/\sqrt 2$, in terms of which operators $C$ and $D$, associated with standing wave modes, can be defined:
\begin{align}
\hat C(t,z) &= \sqrt{\frac 2 \pi} \int e^{-i\omega t}\cos[(k_F +\omega/c)z] \hat c_\omega \, d\omega \cr
\hat D(t,z) &= -i\sqrt{\frac 2 \pi} \int e^{-i\omega t}\sin[(k_F +\omega/c)z] \hat d_\omega \, d\omega 
\label{e5}
\end{align}  
Using these operators, the atom-field interaction Hamiltonian can be written as
\begin{align}
H_I = &\hbar g e^{-i\delta t}\hat C\left(t,\tfrac{a}{2}\right)\bigl(\ket +  \bra{gg} + \ket{ee} \bra{+} \bigr) \cr
&+\hbar g e^{-i\delta t} \hat D\left(t,\tfrac{a}{2}\right)\bigl(\ket -  \bra{gg} - \ket{ee} \bra{-} \bigr) + H.c.
\label{e6}
\end{align}
where we have introduced the collective ``bright'' and ``dark'' states $\ket{\pm} = (\ket{eg}\pm\ket{ge})/\sqrt 2$.  The state $\ket +$ (resp. $\ket -$) couples only to photons in the $C$ (resp. $D$) wave mode.  A further simplification is that the atomic interaction Hamiltonian (\ref{e4}) is  diagonal in this basis:

\begin{equation}
H_A = \hbar\Delta \bigl(\ket{+}\bra{+} - \ket{-}\bra{-}\bigr) +  \hbar\beta\ket{ee}\bra{ee}
\label{e7}
\end{equation}
Apart from their mathematical convenience, the $\hat c$, $\hat d$ modes can also be excited directly with an arrangement as shown in Fig.~2.  An incoming traveling-wave mode with operator $\hat c_{in}$ is transformed by the beamsplitter into the superposition $(\hat a+\hat b)/\sqrt 2$; when the $\hat a$ and $\hat b$ traveling waves arrive back at the beamsplitter, they are in turn transformed into the superposition $\hat c_{out} = (a+b)/\sqrt 2$, which travels in the opposite direction to $\hat c_{in}$.  A similar retroreflection happens to an incoming $\hat d$ field, except that here we have assumed (to preserve unitarity) that $\hat d_{out} = -\hat d_{in}$.

\begin{figure}
\includegraphics[height=5.3cm]{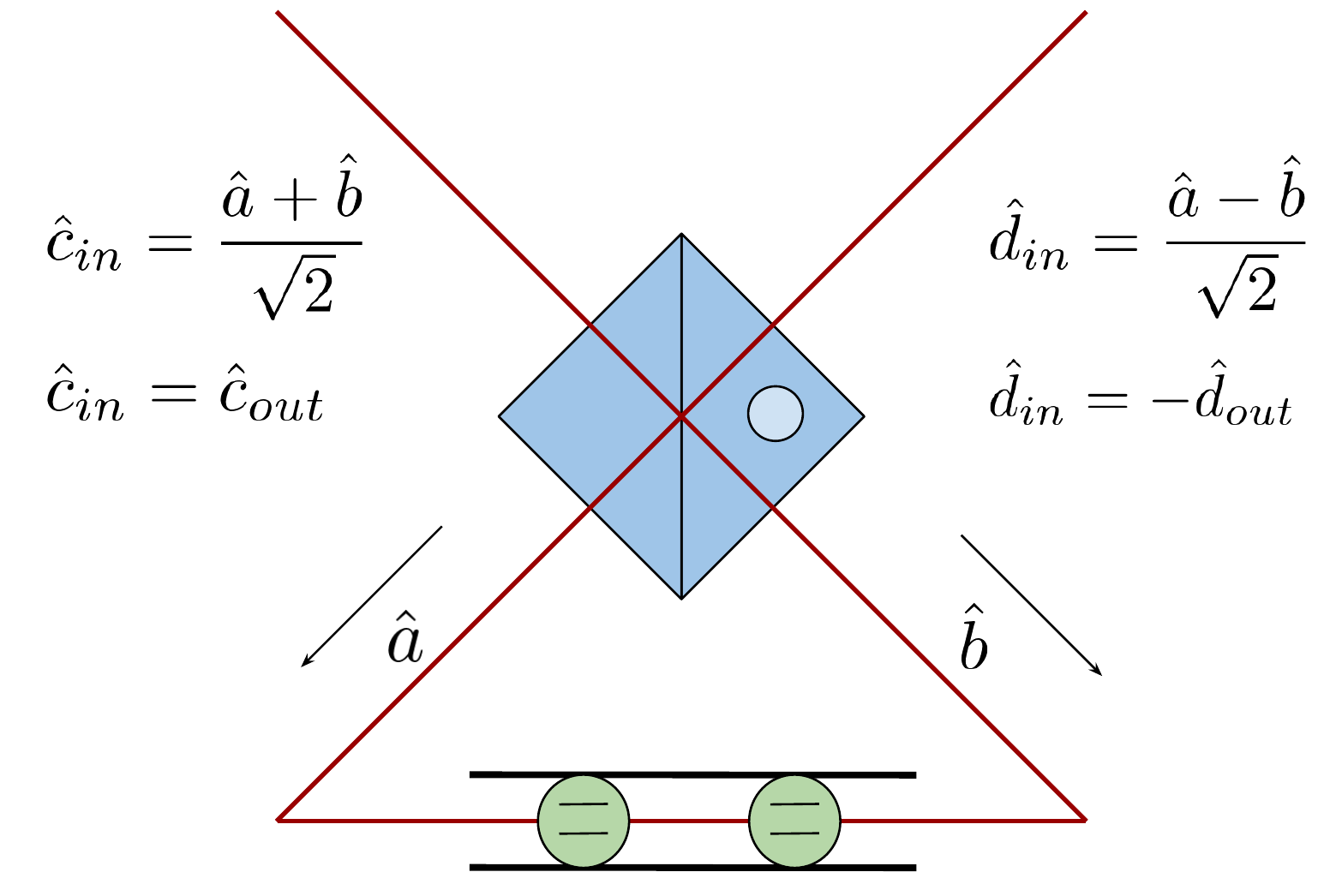}
\caption{\label{fig2} How to excite the standing-wave modes in the waveguide. Incoming running waves in either mode $\hat c$ or $\hat d$ are transformed by the beam splitter into standing waves associated with the operators $(\hat a + \hat b)/\sqrt 2$ and $(\hat a - \hat b)/\sqrt 2$, respectively.  In the absence of interaction, each incoming mode will come back upon itself at the beamsplitter, with an overall phase difference of $\pi$ between them.}
\end{figure}

Since all these fields (the incoming and outgoing traveling waves, and the corresponding standing wave in the waveguide) have the same spectrum, we will refer to all of them as simply the ``$c$'' (or ``$d$'') field, letting the context make it clear which one we have in mind.

%Next: work out the solution for the $c$-type mode, in as few equations as possible$\ldots$
\subsection{Standing-wave solution}

Consider, for definiteness, the case that starts with the atoms in the ground state and a single photon in the $c$ mode.  The  system state at any time can be written as $\ket{\Psi(t)} = \ket{\psi_{g}(t)}\ket{gg} + \ket{\psi_{+}(t)}\ket{+}$, where $\ket{\psi_{g}(t)}$ is a single-photon field state and $\ket{\psi_{+}(t)} \equiv \psi_+(t)\ket 0$.  In a new interaction picture, and with the definition $\delta_\pm = \delta \mp \Delta$, the Schr\" odinger equation becomes
\begin{subequations}
\begin{align}
\frac{d}{dt}\ket{\psi_{g}(t)} &= -ig e^{i\delta_+ t}\hat C^\dagger\left(t,\tfrac{a}{2}\right) \ket{\psi_{+}(t)} \label{e8a} \\
\frac{d}{dt}\ket{\psi_{+}(t)} &= -ig e^{-i\delta_+ t}\hat C\left(t,\tfrac{a}{2}\right)  \ket{\psi_{g}(t)} \label{e8b} 
\end{align}
\label{e8}
\end{subequations}
As we did in \cite{us}, we can formally integrate Eq.~(\ref{e8a}) and substitute in (\ref{e8b}), then put the resulting equation in normal order. The commutation relations for $\hat C,\hat C^\dagger$ are
\begin{align}
\left[\hat C\left(t,\tfrac{a}{2}\right),\hat C^\dagger\left(t^\prime,\tfrac{a}{2}\right)\right] = &2\delta\left(t-t^\prime\right) +e^{i k_F a} \delta\left(t-t^\prime-a/c\right) \cr
&+e^{-i k_F a}\delta\left(t-t^\prime+a/c\right) 
\label{e9}
\end{align}
leading to the following equation for $\ket{\psi_{+}(t)}$
\begin{align}
\frac{d}{dt}\ket{\psi_{+}(t)} = &-g^2 \ket{\psi_{+}(t)} - g^2 e^{ik_F a-\delta_+ a/c} \ket{\psi_{+}(t-a/c)} \cr
&-ig e^{-i\delta_+ t} \hat C\left(t,\tfrac{a}{2}\right)\ket{\psi_{g}(0)}
\label{e10}
\end{align}
Equation (\ref{e10}) already displays the main difficulty one encounters in the multi-atom problem: namely, it is a differential-difference equation, and as such it is virtually impossible to solve analytically in closed form.  It is this that will make it necessary to resort to the ``Markov approximation'' in order to get manageable expressions in the two-photon case.

In this case, however, if we are only interested in the asymptotic state of the field for long times, we can actually obtain a closed-form expression as follows.  Integrating Eq.~(\ref{e8a}) from $t=0$ (by which we mean a time long before the pulse arrives and the atoms are excited, so this lower limit can be formally extended to $-\infty$) to $t=\infty$, we get 
\begin{align}
\ket{\psi_{g}(\infty)} = &\ket{\psi_{g}(0)}-ig \sqrt{\frac 2 \pi} \int_{-\infty}^\infty d\omega\, \cos[(k_F +\omega/c)a/2] \cr
&\qquad\times\left( \int_{-\infty}^\infty e^{i(\delta_+ +\omega)t} \psi_+(t)\, dt\right)\, \hat c_\omega^\dagger \ket 0
\label{e11}
\end{align}
which shows that in order to get the asymptotic spectrum of the field we only need the Fourier transform of the function $\psi_+(t)$, which can be obtained straightforwardly from (\ref{e10}) (it is in inverting the Fourier transform where the difficulty lies).  The final result is then that $\ket{\psi_{g}(\infty)} = \int d\omega\, \tilde f_g^{(c)}(\omega,\infty) \hat c^\dagger \ket 0$, with
\begin{equation}
\tilde f_c(\omega,\infty) = -\frac{g^2+g^2 e^{-i(k_F+\omega/c)a} +i(\omega+\delta_+)}{g^2+g^2 e^{i(k_F+\omega/c)a} -i(\omega+\delta_+)}\tilde f_c(\omega,0)
\label{e12}
\end{equation}
where the function $\tilde f_c(\omega,0)$ is the initial field spectrum.  The corresponding result if one starts with the $d$ modes  instead is
\begin{equation}
\tilde f_d(\omega,\infty) = -\frac{g^2-g^2 e^{-i(k_F+\omega/c)a} +i(\omega+\delta_-)}{g^2-g^2 e^{i(k_F+\omega/c)a} -i(\omega+\delta_-)}\tilde f_d(\omega,0)
\label{e13}
\end{equation}
%Equations (\ref{e12}) and (\ref{e13}) are reminiscent of the single-atom, single-photon result 
The $\omega a/c$ terms in the exponents of Eqs.~(\ref{e12}) and (\ref{e13}) are the ``non-Markovian'' terms.  Their size is of the order of $(a/c)/T$, where $T$ is the duration of the incident pulse, and $a/c$ is the time needed for the field to travel from one atom to the other.  Normally one would expect such terms to be extremely small.  Ignoring them for a moment, we see that Eqs.~(\ref{e12}) and (\ref{e13}) take the same form as the result for a single-atom scattering in a unidirectional waveguide, except for the fact that the coupling is modified, from $g^2$ to $g^2(1\pm\cos(k_F a))$, and the appearance of an additional detuning, $\mp g^2 \sin(k_F a)$.  

The change in coupling is easy to understand:  the atoms are located in a standing wave where the field intensity naturally changes from place to place.  The case $k_F a = 2\pi n$ corresponds to $a=n\lambda$, in which case, with the symmetrical arrangement assumed, the atoms are sitting at the nodes of the $d$-type standing wave and the antinodes of the $c$-wave.  Conversely, maximum coupling to the $d$-type wave happens when $a = \lambda/2 + n\lambda$.  Note, however, that there is also a collective aspect to the increased coupling: in the maximum case, the coupling will go as $2g^2$, twice the size of the result for a single atom at a standing-wave antinode. This is basically a   superradiance effect \cite{Dicke, Haroche}.

The origin of the position-dependent detuning, on the other hand, is less transparent in the standing-wave picture, so we turn to a traveling-wave description next.

\subsection{The traveling-wave solution}

\subsubsection{Non-interacting atoms and cavity analogy}
If the initial pulse is a traveling wave, such as a superposition of (right-traveling) $a$-modes, we can write its initial state as
\begin{align}
\ket{\psi_g(0)} &= \int d\omega\, \tilde f_a(\omega,0) \hat a^\dagger_\omega \ket 0 \cr
&= \frac{1}{\sqrt 2}\int d\omega\, \tilde f_a(\omega,0) \left(\hat c^\dagger_\omega + \hat d^\dagger_\omega\right)\ket 0
\label{e14}
\end{align}
and so we could obtain the final spectrum for the transmitted field simply by adding Eqs.~(\ref{e12}) and (\ref{e13}), with $\tilde f_c(\omega,0) = \tilde f_d(\omega,0) = \tilde f_a(\omega,0)/\sqrt 2$ and dividing by $\sqrt 2$.  The result (in the non-interacting atom case, $\Delta =0$) is
\begin{equation}
\tilde f_a(\omega,\infty) = -\frac{(\omega+\delta)^2}{(g^2-i(\omega+\delta))^2-g^4e^{2i\phi(\omega)}}\tilde f_a(\omega,0)
\label{e15}
\end{equation}
where, to simplify the expressions, we have defined $\phi(\omega) \equiv (k_F+\omega/c)a$. Similarly, subtracting Eqs.~(\ref{e13}) from (\ref{e12}) yields for the reflected field
\begin{align}
\tilde f_b(\omega,&\infty) = g^2 e^{-i\phi(\omega)} \tilde f_a(\omega,0) \cr
&\times \frac{(1+e^{2 i\phi(\omega)})(g^2-i(\omega+\delta))-2g^2 e^{2i\phi(\omega)}}{(g^2-i(\omega+\delta))^2-g^4e^{2i\phi(\omega)}}
\label{e16}
\end{align}
Inspection shows Eqs.~(\ref{e15}) and (\ref{e16}) to be identical to the corresponding Eqs.~(37) and (38) in \cite{zubairy1}, except for a difference in the overall sign of Eq.~(\ref{e16}) for which we have not been able to account. Equation (\ref{e15}) has also been presented in \cite{zheng2}, in a form that also allows for external losses (see also \cite{garciavidal1}).

Interestingly, the above results can be derived in a very straightforward manner by treating each atom as a linear (but frequency-dependent) beamsplitter and adding successive reflections and transmissions as one would in classical optics (for instance, in the study of multiple reflections from a dielectric slab, or a Fabry-Perot cavity).  This is also, essentially, the idea behind the transfer-matrix approach to this problem, which has been presented for $N$ scatterers in \cite{shen2,tsoilaw}.    In our formalism, we find the only slightly nontrivial part is to keep track of the phase factors experienced by the fields upon reflection.  A right-traveling field reflected by an atom at position $z_0$ sees its spectrum modified by 
\begin{equation}
\tilde f_b(\omega) = -\frac{g^2 e^{2i(k_F+\omega/c)z_0}}{g^2-i(\omega+\delta)}\tilde f_a(\omega) \equiv r_R(z_0)\tilde f_a(\omega)
\label{e17}
\end{equation}
This phase shift takes into account the change in position of the pulse (which depends on the position of the atom) compared to a reference left-traveling wave (compare with Eq.~(30) of \cite{zubairy1}).  Similarly, a left-traveling field reflected by an atom at position $z_0$ sees its spectrum modified by 
\begin{equation}
\tilde f_a(\omega)  = -\frac{g^2 e^{-2i(k_F+\omega/c)z_0}}{g^2-i(\omega+\delta)}\tilde f_b(\omega)  \equiv r_L(z_0)\tilde f_b(\omega) 
\label{e18}
\end{equation}
A transmitted field, on the other hand, merely has its spectrum multiplied by a transmission coefficient $t$ given by 
\begin{equation}
t = -\frac{i(\omega+\delta)}{g^2-i(\omega+\delta)}
\label{e19}
\end{equation}
In our case, then, we can define for the atom located at $z_0 = -a/2$ a reflection coefficient $r= r_R(-a/2)$ for a right-traveling wave, and a coefficient $r^\prime = r_L(-a/2)$ for a left-traveling wave and just note that the roles of $r$ and $r^\prime$ are reversed for the atom at $z_0 = a/2$.  For a wave incident from the left, then, we obtain the overall reflection coefficient
\begin{equation}
r+t^2 r^\prime + t^2{r^\prime}^3 + \ldots = r + \frac{t^2 r^\prime}{1-{r^\prime}^2}
\label{e20}
\end{equation}
It is straightforward to verify that the result of this expression is precisely the reflection coefficient shown in Eq.~(\ref{e16}) above.  Similarly, the overall transmission coefficient resulting from multiple reflections inside the two-atom ``cavity'' is
\begin{equation}
t^2 + t^2 {r^\prime}^2 + \ldots =  \frac{t^2}{1-{r^\prime}^2}
\label{e21}
\end{equation}
and this yields precisely Eq.~(\ref{e15}).

The possibility to think of the two atoms as a ``cavity'' explains why changing the atomic separation (contained in the term $\phi(\omega)$ in Eqs.~(\ref{e15}) and (\ref{e16})) affects both the amplitude and the phase of the reflected and transmitted fields---or, equivalently, acts as both an effective detuning and an effective change in the coupling strength.   Modifying the atomic separation changes which frequencies are resonant with the cavity and, correspondingly, the transmitted and reflected spectra.  ``Atomic cavities,'' that is, cavities whose ``mirrors'' are atomic systems (typically consisting of more than one atom each), have been considered for various purposes by a number of authors \cite{KimbleCav,scarani2,zubairy2}.  In the present context, the ``cavity'' formed by just the two atoms, and particularly the stationary states of the field inside, have been studied by Gonzalez-Ballestero, Garcia-Vidal and Moreno in \cite{garciavidal1}.

Results for the reflectivity of the two non-interacting atom system, for an initial Gaussian-shaped pulse, have been presented in \cite{zubairy1}, and can be derived in a straightforward way from the work above, whereas the transmission as a function of frequency for monochromatic fields has been considered in \cite{zheng2,tsoilaw} (see also \cite{scarani1}).  In agreement with these latter works, we find that the frequency-dependent intensity transmission coefficient (absolute value squared of $\tilde f_a(\omega,\infty)/\tilde f_a(\omega,0)$) has the relatively simple form \begin{equation}
{\cal T}(\omega,\phi) = \frac{(\omega+\delta)^4}{(\omega+\delta)^4+4g^4\left((\omega+\delta)\cos\phi+g^2\sin\phi\right)^2} 
\label{e22}
\end{equation}
It is easier to analyze this result by making the Markov approximation, which in this context simply means ignoring the dependance of $\phi$ on $\omega$ (meaning $\phi=k_F a$), so we will do that from now on. Note, however, that including the $\omega a/c$ term to first order in  (\ref{e22}) would still result in an equation of the same form, and hence it would not affect the following considerations substantially.

The most interesting departure Eq.~(\ref{e22}) exhibits with the single-atom case, as has been already noted by other authors, is the opening of transmission ``windows.'' In particular, even though transmission will always be zero at exact resonance, i.e., $\delta =\omega=0$ (just as for a single atom), one can now make ${\cal T}(0,\phi)=1$ for any nonzero $\delta$ and an appropriate value of $\phi$.  Inspection of (\ref{e22}) shows that these transmission resonances happen at $\tan(k_F a) = -\delta/g^2$ \cite{zheng2,tsoilaw,zubairy2}, for which 
\begin{equation}
{\cal T}(\omega,-\tan^{-1}(\delta/g^2)) = \frac{1}{1+4g^4\omega^2/(\omega+\delta)^4}
\label{e23}
\end{equation}
For small enough $\delta/g^2$, these are narrow, slightly asymmetric resonances, of width $\sim \delta^2/g^2$.  For larger $\delta/g^2$, they broaden substantially and become more markedly asymmetric.  This is illustrated in Figure 3, which shows the transmission curves as a function of $\omega/\delta$ and $\phi$ for various values of $\delta/g^2$.

\begin{figure}
\includegraphics[width=8.3cm]{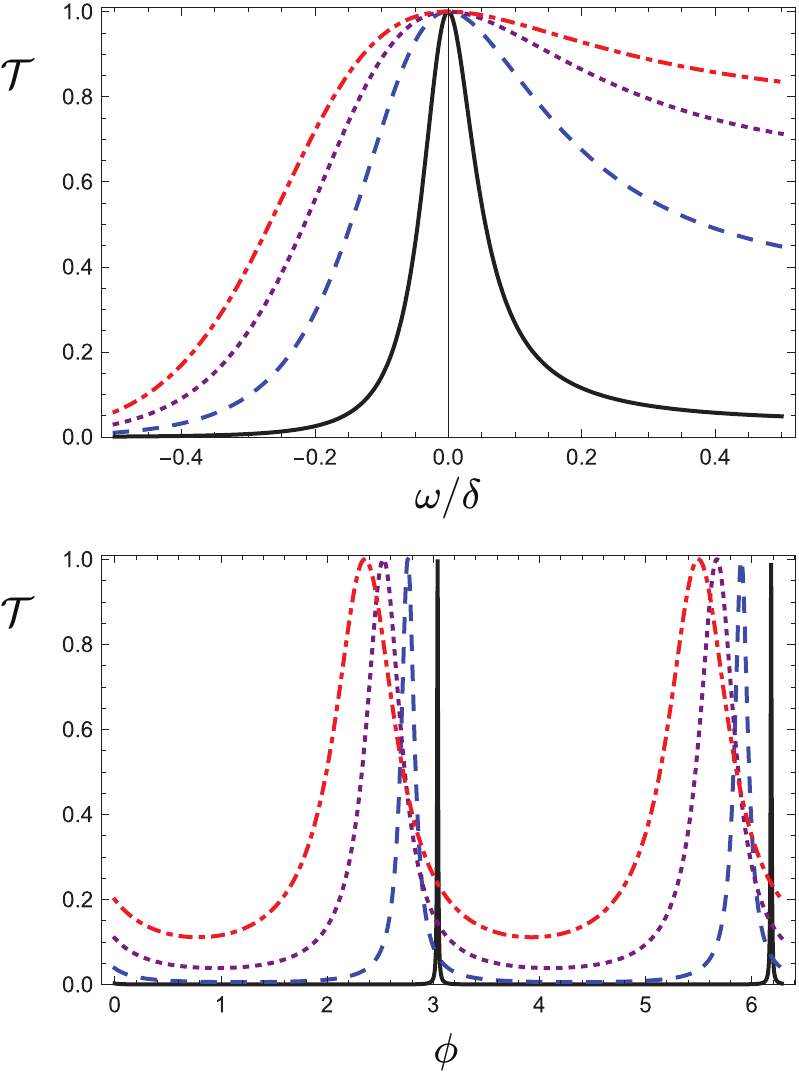}
\caption{\label{fig3} The transmission coefficient ${\cal T}(\omega,\phi)$ for $\delta/g^2 = 0.1$ (solid curve), $0.4$ (dashed), $0.7$ (dotted), and $1$ (dot-dashed).  Top: ${\cal T}$ as a function of $\omega/\delta$ for $\phi = -\tan^{-1}(\delta/g^2)$.  Bottom: ${\cal T}$ as a function of $\phi$ for $\omega=0$. }
\end{figure}

\subsubsection{Interacting atoms}
In the presence of an exchange-type interaction between the two atoms (or other two-level systems), the  spectral amplitude transmission and reflection coefficients corresponding to Eqs.~(\ref{e15}) and (\ref{e16}) can be obtained in exactly the same way from Eqs.~(\ref{e12}) and (\ref{e13}).  For brevity, we present here only the spectral intensity transmission coefficient (corresponding to Eq.~(\ref{e22})):
%\begin{widetext}
%\begin{equation}
%T(\omega,\phi) =\frac{\left(-\Delta ^2-2 \Delta g^2 \sin (\phi )+(\omega+\delta) ^2\right)^2}{\left(-\Delta ^2-2
%   \Delta g^2 \sin (\phi )+(\omega+\delta)^2\right)^2+4 g^4(\Delta +(\omega+\delta)  \cos (\phi )+g^2\sin (\phi
%   ))^2}
%\end{equation}
%\label{e24}
%\end{widetext}
\begin{equation}
{\cal T}(\omega,\phi) =\left(1+\frac{4 g^4(\Delta +(\omega+\delta)  \cos (\phi )+g^2\sin (\phi
   ))^2}{\left(-\Delta ^2-2 \Delta g^2 \sin (\phi )+(\omega+\delta) ^2\right)^2} \right)^{-1}
\label{e24}
\end{equation}
The presence of $\Delta$ in this expression makes a substantial difference with the previous case, as now it is possible to have unit transmission on resonance for any non-zero value of $\Delta$, provided $\phi$ is chosen appropriately.  
The  condition for this is
\begin{equation}
\Delta + \delta\cos\phi + g^2\sin\phi = 0
\label{e25}
\end{equation}
Figure 4 shows some sample transmission curves for $\Delta\ne 0$ and $\delta =0$.  For the curves in Figure 4a, the phase $\phi$ has been chosen to satisfy Eq.~(\ref{e25}).  Figure 4b then shows the sensitivity of the transmission (at $\omega=0$) to the choice of $\phi$.  Note the differences with Figure 3: the transmission curves are now symmetric in $\omega$, and the $\phi$ dependence no longer has period $\pi$, but, rather, $2\pi$.

\begin{figure}
\includegraphics[width=8.3cm]{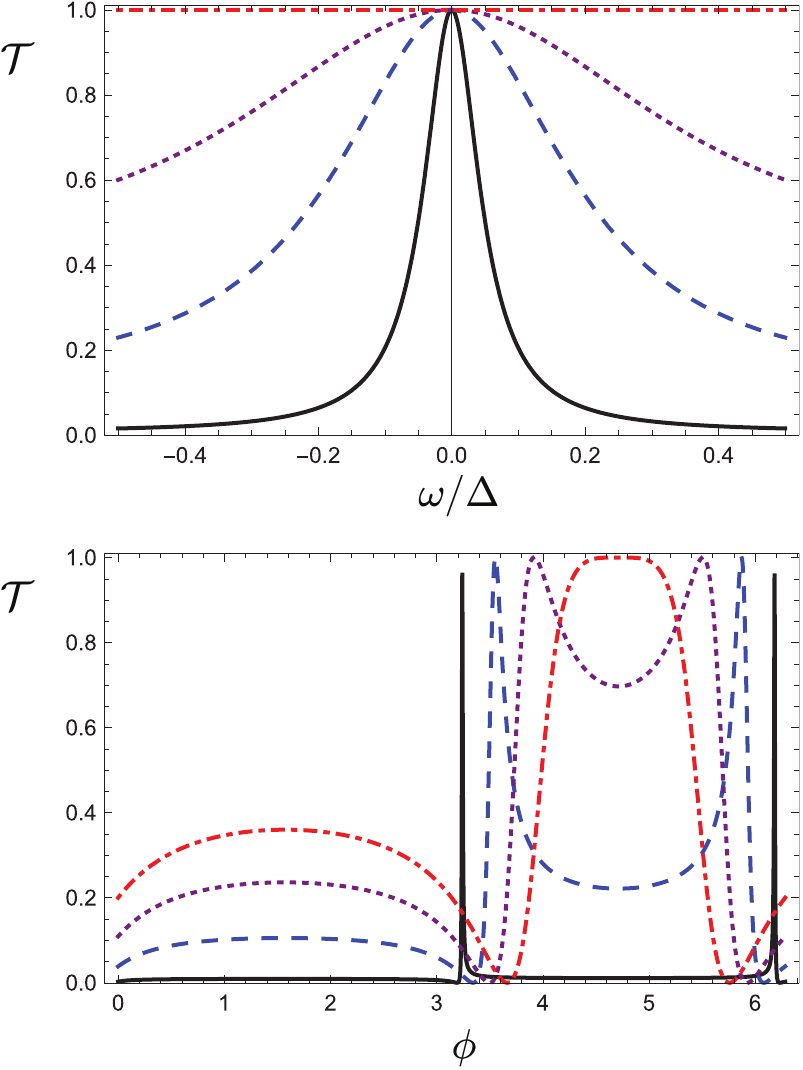}
\caption{\label{fig3} The transmission coefficient ${\cal T}(\omega,\phi)$ for $\delta=0$ and $\Delta/g^2 = 0.1$ (solid curve), $0.4$ (dashed), $0.7$ (dotted), and $1$ (dot-dashed).  Top: ${\cal T}$ as a function of $\omega/\Delta$ for $\phi = -\sin^{-1}(\Delta/g^2)$.  Bottom: ${\cal T}$ as a function of $\phi$ for $\omega=0$. }
\end{figure}

Most remarkable is what happens when $\Delta = g^2$ and $\phi=3\pi/2$ (or $\Delta = -g^2$ and $\phi=\pi/2$, if negative values of $\Delta$ are possible), as per condition (\ref{e25}):  the transmission curve becomes spectrally flat.  The pulse will be entirely transmitted, regardless of its shape, at least to the extent that one can neglect the contribution of the term $\omega a/c$ to $\phi$.  Note also, from Fig.~4b, that the range of $\phi$ over which this holds is remarkably broad.  

This transmission window may actually be viewed as a result of quantum interference.  At $\phi=3\pi/2$ or $\pi/2$, the incoming traveling-wave photon can be expressed as a superposition of standing waves coupling with equal strength to the $\ket +$ and $\ket -$ states. If, at that point, $\Delta$ has the right magnitude and sign to cancel the terms proportional to $g^2e^{i\phi}$ in Eqs.~(\ref{e12}) and (\ref{e13}), the amplitudes for the two processes will be identical and therefore they will cancel exactly when calculating the spectral amplitude of the reflected pulse. For the transmitted pulse, on the other hand, they add to yield 
\begin{equation}
\tilde f_a(\omega,\infty) = -\frac{g^2+i(\omega+\delta)}{g^2-i(\omega+\delta)}\tilde f_a(\omega,0)
\label{ne26}
\end{equation}
This is actually the single-atom, standing-wave (or unidirectional waveguide) result, but here it appears in an explicitly bidirectional setting.  It is a highly nontrivial result, depending as it does on the presence of a second atom and the direct interaction between the atoms. We note here that a transmission window resulting from interaction between the atoms has also been reported by Liao, Nha and Zubairy in \cite{interactions}.

\begin{figure*}
\includegraphics[width=16cm]{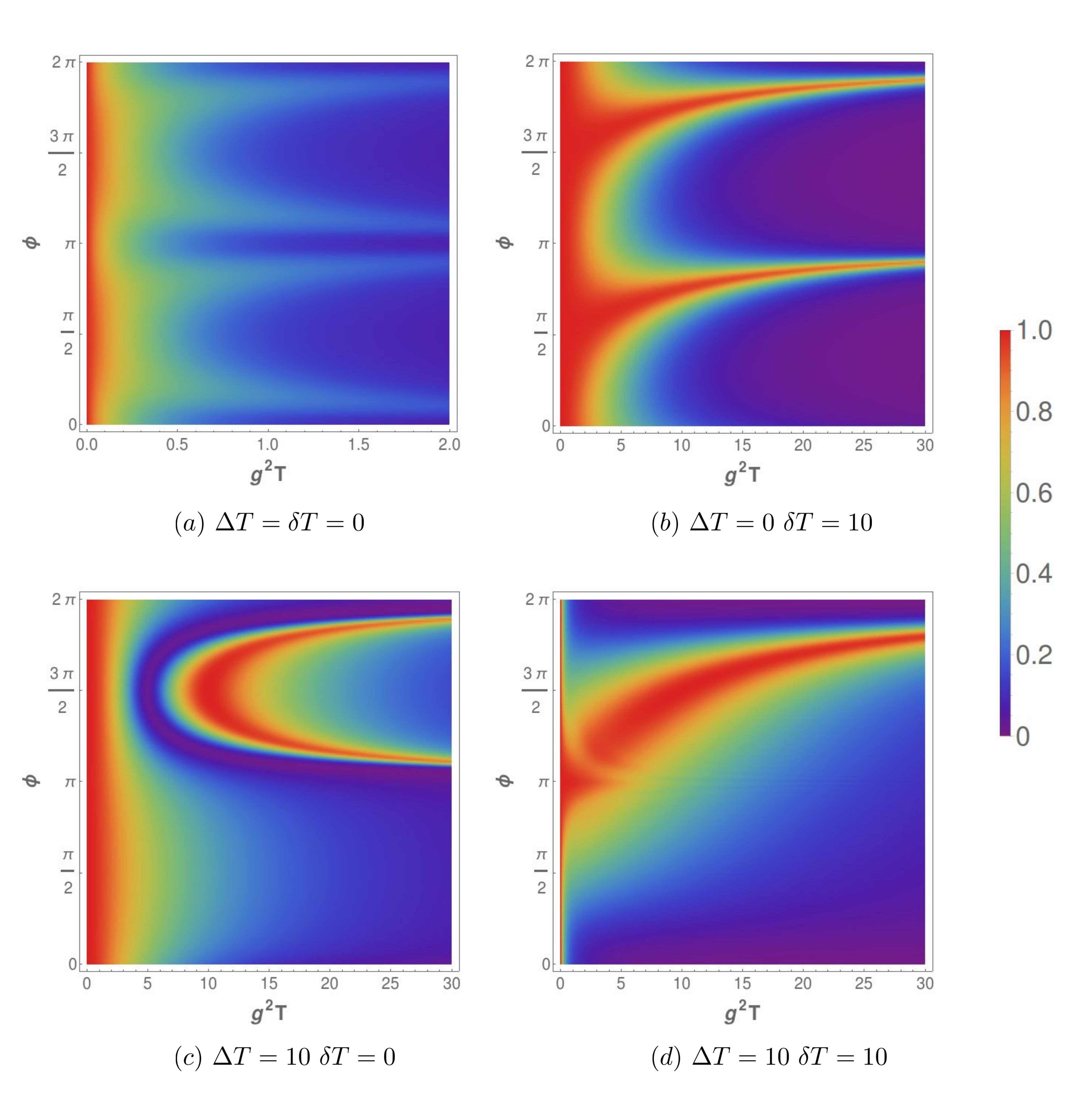}
\caption{\label{fig5} Transmission of a square pulse of duration $T$ through the two-atom ``cavity'' system, as a function of $\phi$ and $g^2 T$, for the values of $\delta T$ and $\Delta T$ shown. }
\end{figure*}

A visual summary of most of the above discussion is provided by Figure 5 below, which shows the intensity transmission coefficient calculated for a square pulse of duration $T$.  Figure 5a shows that for $\delta = \Delta = 0$, one only finds a large transmission for very short pulses, whose frequency spectrum is much broader than the effective width (of the order of $g^2$) of the atomic resonance (alternatively, one can think of the $g^2T\ll 1$ case as the small coupling limit).  Figure 5b is plotted assuming a constant value of $\delta T$, and so, if one imagines $g^2$ to have a fixed value and $T$ to increase towards the left, one must also imagine $\delta$ decreasing accordingly.  The high transmission zones then correspond to the condition $\tan\phi = -\delta/g^2$ discussed above, which for $\delta T =10$ can be written $\phi = -\tan^{-1}(10/g^2T)$.  Similarly, Figure 5c, for $\delta=0$ and $\Delta T = 10$, shows the large transmission zone around $\Delta = g^2$ and $\phi=-3\pi/2$, and also the narrower transmission regions corresponding to condition (\ref{e25}), which here becomes $\phi = -\sin^{-1}(10/g^2T)$. (Note also the high reflection condition $\Delta^2 + 2\Delta g^2\sin\phi=\delta^2$.) 

Finally, Figure 5d gives an idea of what is possible when both $\delta$ and $\Delta$ are nonzero.  Here, note particularly the very large high-reflection region around $\phi=0$, and the high transmission along $\delta/g^2 = -\tan(\phi/2)$.  These are consistent with Eq.~(\ref{e24}), which, for $\delta=\Delta$ and $\omega=0$, predicts ${\cal T} = 1/(1+(\cot(\phi/2) + g^2/\delta)^2)$.

%\subsection{Remarks on the $N$-atom case}

%Enough to show how the solution would go$\ldots$

\section{Two photons, two atoms}

\subsection{Markov approximation and general solution}

The discussion in the previous section showed that, for non-interacting atoms, the single-photon problem is essentially linear: one just has to treat each atom as a frequency-dependent beamsplitter and add the results of multiple, independent, reflections and transmissions.  This changes when one allows for interaction between the atoms, as we have shown.

When the incoming pulse contains two photons, we also expect nonlinear effects to appear on top of the single-photon picture \cite{zheng4}.  Our primary goal here is to identify and characterize the contribution of such (typically entangling) nonlinear processes to the two-atom scattering problem, and to look for situations where they may add some useful capability to the system (or, of course, where the opposite may be true). 

The Hamiltonian for the problem can still be written in terms of Eqs.~(\ref{e5})--(\ref{e7}) as $H_I+H_A$, only now the doubly excited state $\ket{ee}$ can clearly be accessed (except in the very special case where one of the two incoming photons is in the $c$ mode, and the other in the $d$ mode).  This complicates matters, since $\ket{ee}$ can decay via either $\ket +$ or $\ket -$, so in general one has to use the entire basis of atomic states and both the $C$ and $D$ operators.  Writing the full state of the system as 
\begin{equation}
\ket{\Psi(t)} = \ket{\psi_g(t)}\ket{gg} + \ket{\psi_+(t)}\ket{+}+ \ket{\psi_-(t)}\ket{-}+ \ket{\psi_e(t)}\ket{ee}
\label{e26}
\end{equation}
the equations of motion are now
\begin{subequations}
\begin{align}
\frac{d}{dt}\ket{\psi_{g}} &= -ig \left(e^{i\delta_+ t}\hat C^\dagger(t) \ket{\psi_{+}} + e^{i\delta_- t}\hat D^\dagger(t) \ket{\psi_{-}} \right) \label{e27a} \\
\frac{d}{dt}\ket{\psi_{+}} &= -ig \left(e^{-i\delta_+ t}\hat C(t) \ket{\psi_{g}} + e^{i\delta_+^\prime t}\hat C^\dagger(t) \ket{\psi_{e}} \right) \label{e27b} \\
\frac{d}{dt}\ket{\psi_{-}} &= -ig \left(e^{-i\delta_- t}\hat D(t) \ket{\psi_{g}} - e^{i\delta_-^\prime t}\hat D^\dagger(t) \ket{\psi_{e}} \right) \label{e27c} \\
\frac{d}{dt}\ket{\psi_{e}} &= -ig \left(e^{-i\delta_+^\prime t}\hat C(t) \ket{\psi_{+}} - e^{-i\delta_-^\prime t}\hat D(t) \ket{\psi_{-}} \right) \label{e27d}
\end{align} 
\label{e27}
\end{subequations}
where to avoid clutter we have removed the argument $a/2$ from the operators $\hat C$ and $\hat D$.  Recall $\delta_\pm \equiv \delta \mp \Delta $ are the effective detunings appropriate for excitation of the states $\ket{\pm}$ from $\ket{g}$; similarly, we have defined $\delta_\pm^\prime \equiv \delta \pm \Delta - \beta$ as the effective detunings for excitation of the state $\ket{ee}$ from either $\ket +$ or $\ket -$, respectively. The field states $\ket{\psi_\pm}$ now contain one photon each, and the state $\ket{\psi_e}$ is just proportional to the vacuum: $\ket{\psi_e} = \psi_e(t)\ket 0$.

In order to obtain a manageable solution to Eqs.~(\ref{e27}), we find it necessary to make the ``Markov approximation.'' To illustrate the difficulty, consider just the first step in the solution, where we formally integrate Eq.~(\ref{e27a}) and substitute the result in both Eq.~(\ref{e27b}) and Eq.~(\ref{e27c}), and use the commutation relations (\ref{e9}), with $z=z^\prime=a/2$ (and the corresponding ones for $\hat D$, $\hat D^\dagger$), to put the result in normal order.  When this is done, the equation for $d\ket{\psi_+}/dt$ becomes
%which, as indicated in the previous section, basically amounts to neglecting terms of the form $\omega a/c$, where $\omega$ are the frequencies necessary to describe the incident wavepacket's envelope.  In other words, we assume that the incoming pulse is much longer than the time needed for light to travel from one atom to the other.  Normally, this should be an excellent approximation: A pulse created by the decay of an atom will at most be as short as $\sim 10^{-9}$ s, whereas if the atoms are placed not more than a few microns apart (corresponding to optical wavelengths), $a/c \sim 10^{-14}$ s.  
%Details of the various steps of the solution of Eqs.~(\ref{e27}), and the places where the various approximations have been made, can be found in the Appendix.  Here we will only sketch the first few steps.  As before, we can formally integrate Eq.~(\ref{e27a}) and substitute the result in both Eq.~(\ref{e27b}) and Eq.~(\ref{e27c}), and use the commutation relations (\ref{e9}), with $z=z^\prime=a/2$ (and the corresponding ones for $D$, $\hat D^\dagger$), to put the result in normal order.  When this is done, the equation for $\dot{\ket{\psi_+}}$ becomes
\begin{align}
\frac{d}{dt}\ket{\psi_{+}(t)} &= -g^2\ket{\psi_+(t)} - g^2 e^{ik_Fa-i\delta_+a/c}\ket{\psi_+(t-a/c)} \cr
&\;-g^2 \int_{-\infty}^t dt_1 e^{-i\delta_+(t-t_1)}\hat C^\dagger(t_1)\hat C(t)\ket{\psi_+(t_1)} \cr
&+\text{terms involving $\ket{\psi_-}$, $\ket{\psi_e}$ and $\ket{\psi_g(-\infty)}$}
\label{e28}
\end{align}
In order to proceed, we want to replace $\ket{\psi_+(t-a/c)}$ by $\ket{\psi_+(t)}$, at which point the equation can be formally integrated in turn (note, in passing, that a first-order treatment of the $a/c$ shift by means of a Taylor expansion would also be possible, but we will not pursue this path here).  As indicated earlier, this requires that the incident pulse should be much longer than the time needed for light to travel from one atom to the other. Normally, this should be an excellent approximation: A pulse created by the decay of an atom will at most be as short as $\sim 10^{-9}$ s, whereas if the atoms are placed not more than a few microns apart (corresponding to optical wavelengths), $a/c \sim 10^{-14}$ s.  However, one must also consider the fact that the interaction with the atoms also may introduce frequency components as large as the detunings $\delta$, $\Delta$, and $\beta$, and thus we must also require all these to be much smaller than $c/a$.  

At this point, we may invoke the results in Section 2, which suggest that all the system's interesting behavior will happen when the detunings are at most of the order of $g^2$, so it should be enough to restrict consideration to this range and require $g^2a/c \ll 1$.  Again, however, in the optical domain this should be straightforward, since $g^2$ is, essentially, the atomic decay rate or the single-photon Rabi frequency, and it is hard to envision a situation where this would be much greater than 1 GHz or so.  Notice that this means that we can now, consistently, neglect the term $e^{-i\delta_+a/c}$ in Eq.~(\ref{e28}), and just use $\gamma_+ = g^2(1 + e^{ik_Fa})$ as the (complex) decay rate for the state $\ket{\psi_{+}}$.  

A similar treatment yields then the decay rate $\gamma_- = g^2(1 - e^{ik_Fa})$ for the state $\ket{\psi_{-}}$.  We leave the details of the subsequent steps for the Appendix, where we show how, by consistent, repeated use of this approximation, explicit results for $\ket{\psi_\pm(t)}$ and $\ket{\psi_e(t)}$ in terms of the initial state, $\ket{\psi_0} \equiv \ket{\psi_g(-\infty)}$, can be obtained.  While these equations describe the dynamics of the atomic system at an arbitrary time, here we are primarily interested in the asymptotic state of the scattered field, for which all that is needed is the result for $\ket{\psi_g(\infty)}$.  As detailed in the Appendix, we arrive at

\begin{widetext}
\begin{align}
|\psi_{g}(t)\rangle=|\psi_0\rangle &-g^2 \int^{t}_{-\infty} dt_1\int^{t_1}_{-\infty} dt_2 \Big(e^{-\Gamma_{+}(t_1-t_2)} \hat C^{\dagger}(t_1) \hat C(t_2) + e^{-\Gamma_{-}(t_1-t_2)} \hat D^{\dagger}(t_1) \hat D(t_2)\Big) |\psi_0\rangle\cr
 &+ g^4 \int_{-\infty}^{t}dt_1 \int_{-\infty}^{t_1}dt_2 \int_{-\infty}^{t_2}dt_3 \int_{-\infty}^{t_3}dt_4 \cr
 &\qquad\Big[e^{-\Gamma_{+}(t_1-t_2)} \hat C^{\dagger}(t_1) \left(e^{-\Gamma_{+}(t_3-t_4)} \hat C^\dagger(t_3)    \hat C(t_4) + e^{-\Gamma_{-}(t_3-t_4)}  \hat D^{\dagger}(t_3) \hat D(t_4)\right)\hat C(t_2)\cr
   &\qquad + e^{-\Gamma_{-}(t_1-t_2)} \hat D^{\dagger}(t_1) \left(e^{-\Gamma_{+}(t_3-t_4)}   \hat C^{\dagger}(t_3)  \hat C(t_4) + e^{-\Gamma_{-}(t_3-t_4)} \hat D^\dagger(t_3) \hat D(t_4)\right) \hat D(t_2) \Bigr]\cr
   &+ g^4 \int_{-\infty}^{t}dt_1 \int_{-\infty}^{t_1}dt_2 \int_{-\infty}^{t_2}dt_3 \int_{-\infty}^{t_3}dt_4\Big(e^{-\Gamma_{+}(t_1-t_2) } \hat C^{\dagger}(t_1) \hat C^{\dagger }(t_2)-e^{-\Gamma_{-}(t_1-t_2)}\hat D^{\dagger}(t_1) \hat D^{ \dagger}(t_2)\Big)\cr
   &\qquad\times e^{-(2g^2-i( 2\delta-\beta ))(t_2-t_3)} \Big(e^{-\Gamma_{+}(t_3-t_4) } \hat C(t_3) \hat C(t_4)-e^{-\Gamma_{-}(t_3-t_4)}\hat D(t_3) \hat D(t_4) \Big)|\psi_0\rangle
   \label{e29}
\end{align}
for the ground state at time $t$ in terms of the modified (complex) decay rates $\Gamma_\pm = \gamma_\pm-i\delta_\pm = g^2(1\pm e^{ik_Fa})-i(\delta\mp\Delta)$.  Note that (in the Markov approximation) these are the same rates that appear in the single-photon Eqs.~(\ref{e12}) and (\ref{e13}):  For instance, the fraction in Eq.~(\ref{e12}) could simply be written $(\Gamma_+^\ast +i\omega)/(\Gamma_+-i\omega)$.
\end{widetext}
Although Eq.~(\ref{e29}) looks complicated, its structure is actually relatively straightforward, since it contains all the ways one may end up with two photons in the final state if one starts with two photons.  The fact that it is possible to, essentially, rearrange the perturbation series so that it appears to stop after two absorptions is just a consequence of normal ordering, and is essentially the same result as we derived in \cite{us} (and independently derived by Roulet and Scarani in \cite{roulet}) for the single-atom case.  

The first three terms in Eq.~(\ref{e29}) represent scattering events similar to those described in \cite{us} for a single atom. The photons may not interact with the atoms at all ($\ket \psi_0$), only one photon may interact with one of the $\ket \pm$ states (double-integral term), or both photons may interact with either of the $\ket \pm$ states (first four-fold integral term).  This last term results in time- (or, alternatively, frequency-) entanglement, as the interactions are constrained such that the first photon must be absorbed and re-emitted before the second one can be absorbed.

Finally, the second four-fold integral term in Eq.~(\ref{e29}) involves two consecutive absorptions and is therefore directly due to the presence of the doubly excited state.  As we shall see, it yields a purely entangled state, similar in form to the single-atom entanglement term, even though directly stemming from the presence of a second atom.

\subsection{Standing-wave solution}

Consider first the case where both photons are initially in a ``$c$''-type standing wave mode. Over half the terms in (\ref{e29}) then vanish, and there are only two possibilities for the final state: either both photons end up in $c$ modes, or both end up in $d$ modes, the last outcome being made possible by the doubly excited state.  Since, as shown in Fig.~2, it is possible in principle to excite the separate standing wave modes separately, the possibility of getting the photons to leave through the ``wrong'' port of the beamsplitter, as a genuine two-photon, two-atom effect, warrants to spend a little more time on this case, which we will do in the rest of this subsection. 
We note that, on the other hand, the split input case (one photon initially in the $c$ port and one in the $d$ port) will still result in a split output as the only possible outcome, so any genuine two-photon, two-atom effects present in this scenario (if they can be found at all) would have to be more subtle.

Let then the initial state have the form
\begin{equation}
\ket{\psi_0} = \frac{1}{\sqrt 2} \left(\int d\omega  f(\omega) \hat c_\omega^\dagger \right)^2\ket 0
\label{e30}
\end{equation}
Within the Markov approximation (that is, whenever they are multiplied by something that restricts the range of frequencies $\omega$ so that $|\omega a/c|\ll 1$), the $D$ and $C$ operators are approximately equal to
\begin{align}
\hat C(t,\tfrac{a}{2}) &\simeq 2\cos(k_Fa/2) \,\hat\phi_c(t) \cr
\hat D(t,\tfrac{a}{2}) &\simeq -2i\sin(k_Fa/2)\, \hat\phi_d(t)
\end{align}
where $\hat\phi_c(t)$ and $\hat\phi_d(t)$ are the operators that destroy a $c$ or $d$ photon at the instant $t$ (compare Eq.~(\ref{ne3})),  which means that we can practically read out the two-photon wavefunction of the final state, in the time domain, directly from Eq.~(\ref{e29}).  (See our treatment of Eq.~(8) in \cite{us} for details, particularly for how to extract correctly the entangled part of the state.) We define the auxiliary quantities
\begin{align}
\Gamma_c &= Re(\Gamma_+) = 2g^2\cos^2(k_F a/2) = g^2\left(1+\cos(k_F a)\right) \cr
\Gamma_s &= Re(\Gamma_-) = 2g^2\sin^2(k_F a/2) = g^2\left(1-\cos(k_F a)\right)
\end{align}
as well as the function $G(t)$, whose absolute value square is proportional to the single-photon excitation probability (again, see \cite{us}):
\begin{equation}
G_\pm(t)= \int_{-\infty}^{t}dt^{\prime} e^{-\Gamma_\pm(t- t^{\prime})} f(t^{\prime} )
\end{equation}
and a corresponding function for the doubly excited state:
\begin{equation}
\mathcal{E}_\pm(t)=\int_{-\infty}^{t} d t^{\prime}e^{-(2g^2-i( 2\delta-\beta ))(t- t^{\prime})}  f(t^{\prime})G_\pm(t^{\prime})
\label{e34}
\end{equation}
(for its precise relation to the double-excitation probability, see Eq.~(\ref{a10}) in the Appendix). In terms of these, we find the two-photon wavefunction component corresponding to the ``$c$'' exit channel to have the form
\begin{align}
f_{cc}(t_1,t_2)= &\left(f(t_1)-2 \Gamma_{c} G_+(t_1)\right)\left(f(t_2)-2 \Gamma_{c} G_+(t_2)\right)\cr
&-4  \Gamma_{c}^2 e^{-\Gamma_{+} |t_1-t_2|}\left(G_+^2(t_<)-\mathcal{E}_+(t_<) \right)
\label{e35}
\end{align}
where $t_<$ is the smaller of $t_1$ and $t_2$.  The first term in (\ref{e35}) corresponds to two independent single-photon interactions; the second term is an entangled state, whose first half is the familiar single-atom result from \cite{us}, and whose second part comes from the doubly excited state.

Similarly, the two-photon wavefunction component corresponding to the ``$d$'' exit channel is the pure two-photon, two-atom term
\begin{equation}
f_{dd}(t_1,t_2)= 4\Gamma_c\Gamma_s e^{-\Gamma_{-} |t_1-t_2|}  \mathcal{E}_+(t_<)
\label{e36}
\end{equation}
We thus have a situation where two uncorrelated photons can go in through one port of the beamsplitter, and sometimes come out the other port in an entangled state (whereas a single photon would always leave through the same port it went in).  It would be nice if we could get this to happen all the time (to use it either as a deterministic entangled-state generator, or photon number discriminator), but we find it is generally impossible: the best we have been able to obtain numerically for the norm of the term (\ref{e36}) is approximately  $0.593$ for a square pulse. This happens near $g^2 T = 0.91$, $k_F a=3\pi/2$ and $\Delta=g^2$.
This situation is very similar to the two-photon discriminator suggested by Witthaut and co-workers \cite{witthaut}, which we analyzed in \cite{us}. In fact, when using the same square pulse shape, the device in \cite{witthaut} achieves a nearly identical single-pass maximum success probability of $0.584$ to separate a two photon state from a single photon state.  

Although in general evaluation of the function $\mathcal{E}(t)$ analytically is only possible for special pulse shapes, we can gain some insight by considering the adiabatic limit of a very long (or nearly monocromatic) pulse.  In that limit, as shown in \cite{us}, we have $G_+ \simeq f(t)/\Gamma_+$, and accordingly
\begin{equation}
\mathcal{E}_+(t) \simeq \frac{1}{(2g^2-i( 2\delta-\beta ))(g^2+g^2 e^{ik_F a}-i(\delta-\Delta))} f^2(t) 
\label{e37}
\end{equation}
The two factors in the denominator of this equation represent conditions for 1- and 2-photon resonance.  If we choose  $k_F a= 3\pi/2$ , which maximizes the prefactor $\Gamma_c\Gamma_s$, and also makes $|\Gamma_+|=|\Gamma_-|$, we get the conditions $ \beta=2\delta $ and $g^2=\delta+\Delta$ to maximize the norm of $\mathcal{E}_+$.  Note that it is impossible to satisfy both conditions simultaneously in the absence of atom interactions, that is, one needs either $\beta$ or $\Delta$ to be nonzero.  Under these conditions, the entangled state (\ref{e36}) becomes approximately $2e^{-g^2|t_1-t_2|}f^2(t_<)$ times a phase factor.  This is about a factor of 2 smaller in magnitude than the $G_+^2$ term (which characterizes the single-atom entanglement), for the same choice of parameters.

Other choices, however, can make the two terms comparable. In particular, $k_Fa= 2n\pi$ makes $\Gamma_s=0$ (so Eq.~(\ref{e36}) vanishes), but it also makes it possible to set $\Gamma_+ = 2 g^2$, in which case one can get the whole entangled state in Eq.~(\ref{e35}) to vanish approximately.  (The same can be achieved by choosing $k_Fa= (2n+1) \pi$ and $\Gamma_c=0$.) Besides the condition on $\phi$, this requires a long pulse (so $\omega$ will be small, and the approximation (\ref{e37}) will work), $ \beta=2\delta $ and $\delta=\Delta$, both of which could be satisfied in the absence of interactions by just setting $\delta=0$.  Note that even though in this regime the atoms are interacting with only a single standing-wave mode, there are still two photons present, and so the vanishing of the nonlinear terms is a nontrivial result:  it means one can use the two-atom nonlinearity to cancel out the single-atom one.  

\subsection{Running wave configuration}

\subsubsection{Two photons arriving from the same direction}
For two uncorrelated photons arriving from, say, the $a$ direction, the initial state will be given by Eq.~(\ref{ne3}) with $f(\omega_1,\omega_2) = f(\omega_1)f(\omega_2)$.  In the Markov approximation, as we indicated above, we can approximately rewrite the $C$ and $D$ operators appearing in (\ref{e29}) in terms of $\hat\phi_a$ and $\hat\phi_b$:
\begin{align}
\hat C(t) &= \sqrt 2 \cos(k_F a/2)  \left(\hat\phi_a(t)+\hat\phi_b(t)\right) \cr
\hat D(t) &= -i{\sqrt 2} \sin(k_F a/2) \left(\hat\phi_a(t)-\hat\phi_b(t)\right)
\end{align}
We can use this in Eq.~(\ref{e29}) to select the terms that correspond to the input and output states we are interested in.  For instance, if the incoming state has both photons in $a$-modes, as above, there is no need to consider the contribution of the $\hat\phi_b$ operators to the annihilation operators in (\ref{e29}).  If we are interested in the case where both photons are reflected, then there is no need to consider the contribution of the $\hat\phi_a^\dagger$ operators to the creation operators, and so forth.  Altogether there are three possible outcomes (both photons transmitted, both photons reflected, and the ``split'' case, where one photon is transmitted and the other reflected), whose contributions to the overall wavefunction can be written most compactly in terms of the single-photon transmission and reflection functions
\begin{align}
\tau(t) &= f(t)-\Gamma_c G_+(t) -\Gamma_s G_-(t) \cr
\rho(t) &= -\Gamma_c G_+(t) +\Gamma_s G_-(t) 
\label{ne41}
\end{align} 
as 
\begin{subequations}
\begin{align}
f_{aa}(t_1,t_2) = &\tau(t_1)\tau(t_2) +f_{G^2,aa}^{+}(t_1,t_2)+f_{\mathcal{E},aa}^{+}(t_1,t_2) 
\label{e40a} \\
f_{bb}(t_1,t_2) = & \rho(t_1)\rho(t_2)+f_{G^2,aa}^{-}(t_1,t_2)+ f_{\mathcal{E},aa}^{ + }(t_1,t_2) \label{e40b} \\
f_{ab}(t_1,t_2) = &\tau(t_1)\rho(t_2)+ f_{G^2,ab}(t_1,t_2)+f_{\mathcal{E},aa}^{-}(t_1,t_2)
\label{e40c}
\end{align}
\label{e40}
\end{subequations}
where the nonlinear parts, $f_{G^2,aa}^{\pm}$, $f_{G^2,ab}$, and $f_{\mathcal{E},aa}^{+}$ are given by

\begin{widetext}
\begin{align}
f_{\mathcal{E},aa}^{\pm}(t_1,t_2) =\left(\Gamma_c e^{-\Gamma_+|t_1-t_2|}\pm \Gamma_s e^{-\Gamma_-|t_1-t_2|}\right) \left(\Gamma_c \mathcal{E}_+(t_<)+ \Gamma_s \mathcal{E}_-(t_<)\right)\cr
f_{G^2,aa}^{\pm}(t_1,t_2)=-\left(\Gamma_c G_+(t_<)\pm \Gamma_s G_-(t_<)\right)\left(\Gamma_c e^{-\Gamma_+|t_1-t_2|}G_+(t_<)\pm \Gamma_s e^{-\Gamma_-|t_1-t_2|}G_-(t_<)\right) \cr
f_{G^2,ab}(t_1,t_2)=-\left(\Gamma_c G_+(t_<)- sgn(t_1-t_2) \Gamma_s G_-(t_<)\right)
\left(\Gamma_c e^{-\Gamma_+|t_1-t_2|}G_+(t_<)+ sgn(t_1-t_2) \Gamma_s e^{-\Gamma_-|t_1-t_2|}G_-(t_<)\right)
\end{align}
\label{ne43}
\end{widetext}
All three outcomes, therefore, show different entangled components.  Generally speaking, we still see the tendency for the components associated with the doubly excited state to counter the single-atom nonlinear effects, and, as pointed out above, this effect is most pronounced around $\phi = n\pi$, when the photons couple only to one or the other of the standing-wave modes (that is, either $\Gamma_c$ or $\Gamma_s$ is zero).  Here it is possible to recover essentially the linear results for a suitable choice of detunings, as shown in the previous subsection.  This may be good news if one generally regards the nonlinear effects as a nuisance (for instance, if entanglement is something to be avoided). 

At the opposite end, relatively large nonlinear effects can be seen when the coupling to both $\ket +$ and $\ket -$ has equal strength ($\phi=(n+\frac 1 2)\pi$).  Here, as discussed above, one can have $G^2$ as large as $2\mathcal{E}$, so no cancellation will typically occur between the single- and two-atom terms.  However, in the running-wave configuration, as Eq.~(\ref{ne43}) shows, it is now possible for some of these terms to cancel separately.  For instance, if $\Gamma_+ = \Gamma_-$, one will have $f_{\mathcal{E},aa}^{-} = f_{G^2,aa}^{-} = f_{G^2,ab}=0$.  This exact cancellation is only possible if $\Delta\ne 0$ (another possible advantage of the atom-atom interactions), but approximate cancellations will still happen if only $\Gamma_c=\Gamma_s$.

The effect of the nonlinear terms is most visible when they cause something to happen that would not normally take place (or would be strongly suppressed) in the linear regime; for instance, along the high transmission regions shown in Fig.~5, where they may contribute strongly to reduce the transmission and increase the reflected and/or the split components.  This also makes sense since those special regions are typically located near resonances, which also enhance the size of the nonlinear terms, as suggested by Eq.~(\ref{e37}) above.    

An example of this can be seen in Fig.~6, which shows the results of calculations of the norm of the transmitted, reflected, and split components of the two-photon wavefunction for a square-pulse incident two-photon state.  The value of the coupling $g$ has been chosen so as to display the nonlinear components to maximum advantage.  For small values of $g^2 T$ the nonlinear effects naturally tend to be small, whereas for large values they are suppressed as the wavefunction is ``squeezed'' in the $t_1-t_2$ direction by the factors of $\exp(-\Gamma|t_1-t_2|)$.  The parameters chosen for Fig.~6 correspond to the high transmission condition Eq,~(\ref{e25}) identified in Section II.C.2, which, in particular, yields perfect transmission in the linear regime when $\phi = 3\pi/2$.  Clearly, the nonlinear terms spoil this effect completely, by introducing a substantial probability for the two photons to be reflected.  (Interestingly, the split outcome is still suppressed, which is consistent with the fact that in this case $\Gamma_+  = \Gamma_-$ and $f_{G^2,ab}+f_{\mathcal{E},aa}^{-} =0$, as indicated above.)

%In fact, the coupling assumed in Fig. 6 is so low that the two-photon reflection and transmission probabilities for a single atom (black lines) are still relatively low, and the most likely outcome is a ``split'' scenario.  The figure thus shows also the strong enhancement of transmission that can happen with two interacting atoms, for the right interaction strength and ``cavity length'' (as discussed in Section 2.B.2), and how nonlinear terms tend to     

\begin{figure}
\includegraphics[width=8cm]{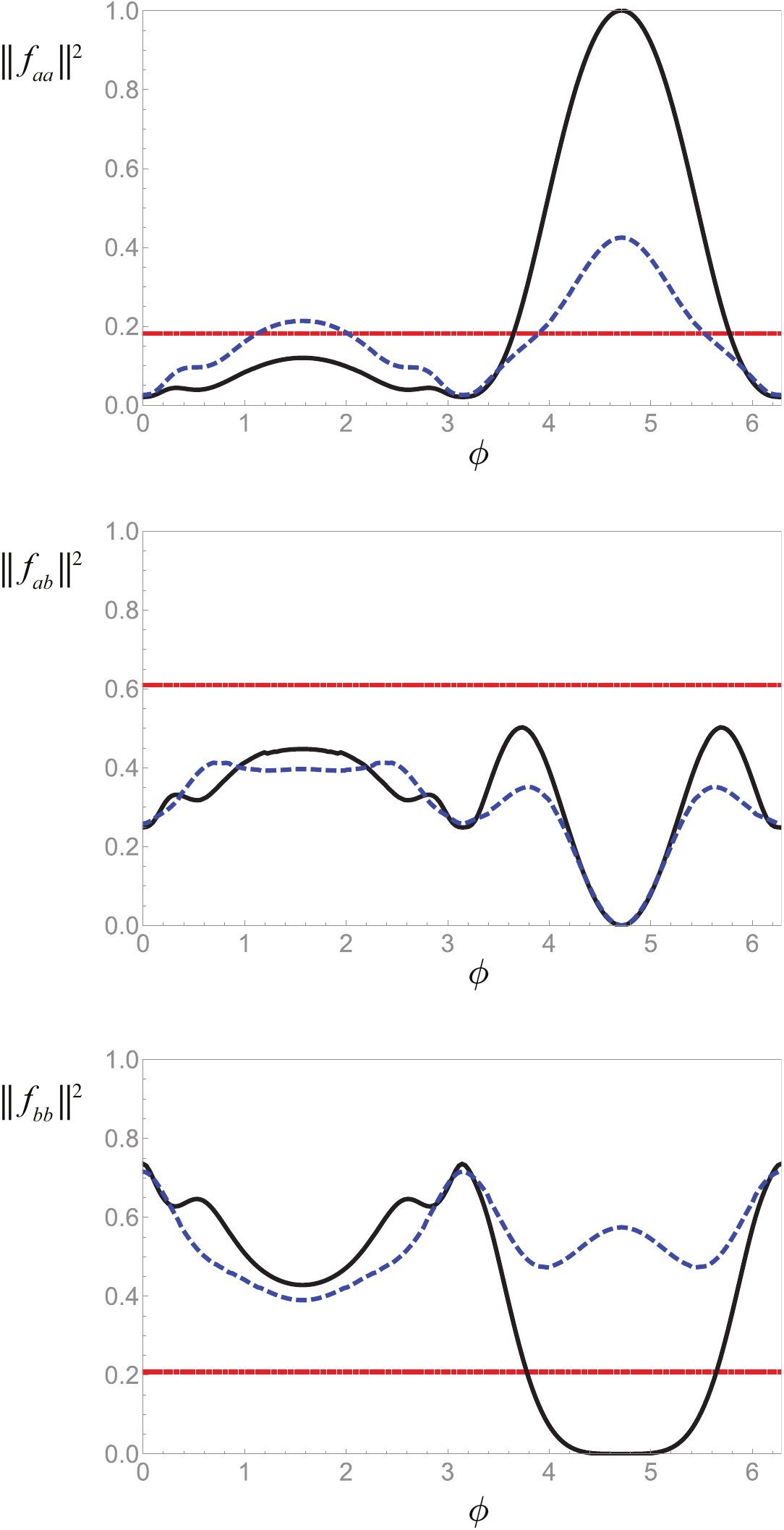}
\caption{\label{fig6} Probabilities of two-photon reflection, one reflection and one transmission, and two-photon transmission when the photons are initially copropagating. Values are calculated for a square pulse of duration $T$, for the parameters $\Delta =g^2 |\sin(k_F a)|$, $\delta =\beta =0$, and $g^2 T=1$.  The dotted (constant) line shows the corresponding probabilities for a single atom. The dashed line shows the result of including all the terms in the wavefunction, whereas the solid line shows only the contribution of the linear terms.}  %
\end{figure}

\subsubsection{Two photons arriving from opposite directions}
When each photon arrives from a different direction, the initial state is 
\begin{equation}
\ket{\psi_0} = \int \int d\omega_1\,d\omega_2 f(\omega_1)f(\omega_2) a^\dagger_{\omega_1} b^\dagger_{\omega_2} \ket 0 \end{equation}
The calculation is a bit more involved in this case, but still straightforward.  The final result can be written as
\begin{align}
f_{aa}(t_1,t_2) &= f_{bb}(t_1,t_2) \cr
&=\frac{1}{\sqrt 2} \Bigl[\tau(t_1)\rho(t_2)+\tau(t_2)\rho(t_1)\Bigr]+\sqrt 2 f_{ent,ab}^+(t_1,t_2) \cr
f_{ab}(t_1,t_2) &= \Bigl[\tau(t_1)\tau(t_2)+\rho(t_1)\rho(t_2) \Bigr]+2 f_{ent,ab}^-(t_1,t_2)\cr
\label{eq45}
\end{align}
in terms of the reflection and transmission functions defined in Eqs.~(\ref{ne41}), and yet a different nonlinear term, this time given by
\begin{align}
f_{ent,ab}^\pm = &-\left(\Gamma_c^2 e^{-\Gamma_+|t_1-t_2|} G_+(t_<)^2\mp \Gamma_s^2 e^{-\Gamma_-|t_1-t_2|} G_-(t_<)^2\right) \cr
&+\left(\Gamma_c e^{-\Gamma_+|t_1-t_2|}\pm \Gamma_s e^{-\Gamma_-|t_1-t_2|}\right) \cr
&\quad\times\left(\Gamma_c \mathcal{E}_+(t_<)- \Gamma_s \mathcal{E}_-(t_<)\right)
\label{ne46}
\end{align}
If we consider the linear terms first, we will see that the probability of a split outcome will approach one whenever the corresponding single-photon case exhibits either unit transmission or unit reflection.  As we saw in Section II.C.2, and demonstrated in Fig.~5, both of these conditions can be fairly accurately predicted, even for finite-bandwidth pulses, by setting $\omega=0$ in Eq.~(\ref{e24}), the results being the conditions (\ref{e25}) for high transmission and $\Delta^2 + 2\Delta g^2\sin\phi=\delta^2$ for high reflection.

Most interesting, of course, is the case of equal coupling $\phi=3\pi/2$, $\Delta=g^2$ where, in the absence of nonlinear effects, both counterpropagating photons would be perfectly transmitted, regardless of the pulse shape.  Recall that for this case $\Gamma_+ =\Gamma_- = \Gamma_c = \Gamma_s = g^2$, and hence the function $f^+_{ent,ab}$ vanishes identically.  Thus, remarkably, the photons still pass through with unit probability, as if they had not interacted with each other, as seen in Fig.~7; yet the function $f^-_{ent,ab}$ is non zero, so there is a nonlinear contribution to the wavepacket distortion and the final state is at least partially entangled.  Specifically, we have
\begin{align}
f_{ab}(t_1,t_2) = &(f(t_1)-2g^2 G(t_1))(f(t_2)-2g^2 G(t_2)) \cr
&- 4 g^2 e^{-g^2|t_1-t_2|}G(t_<)^2
\label{ne47}
\end{align}
The result (\ref{ne47}) is actually very interesting.  A direct comparison with \cite{us} shows that it is identical to the two-photon wavefunction resulting from the scattering of two photons from a single two-level emitter in a unidirectional (or standing wave) geometry; however, here it shows up in a situation in which the two photons are traveling in opposite directions and can then be easily separated before and after the interaction.  Further, a look at the form of (\ref{ne47}) in frequency space (see, for instance, Eq.~(A4) of \cite{us}) shows that this is also the same transformation experienced by a two-photon wavepacket incident on a three-level atom in the ``V'' configuration, each one coupled to one of the two transitions (compare, for instance, \cite{chudziki,brod2}).  This last fact is not surprising, since the level structure of the two-atom system in this regime is isomorphic to a single V system (the transitions in question being $\ket{gg}\to \ket +$ and  $\ket{gg}\to \ket -$).

\begin{figure}
\includegraphics[width=8cm]{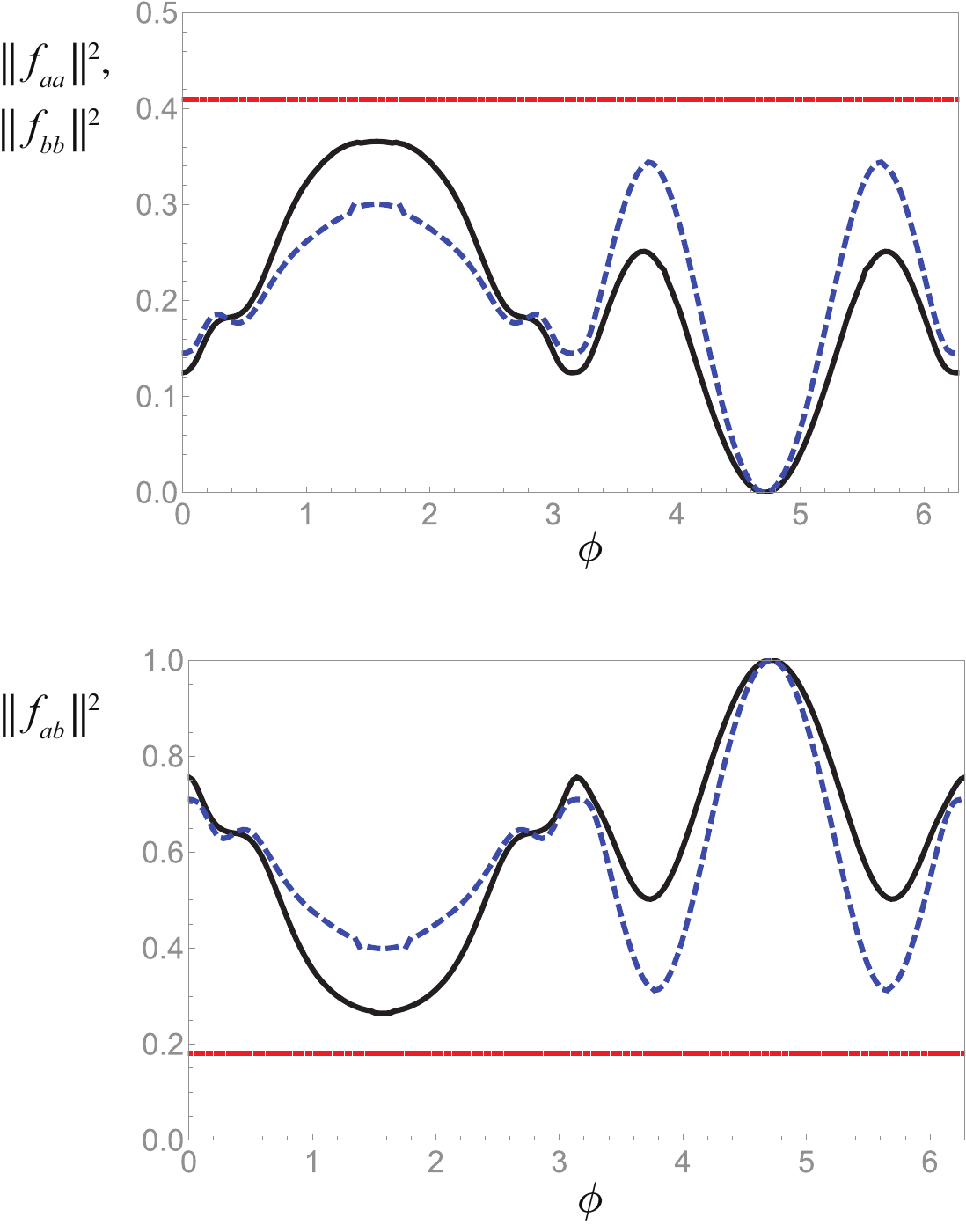}
\caption{\label{fig7} Probabilities of both photons leaving in the same direction (top) or in opposite directions (bottom), when the photons are initially counterpropagating. Values are calculated for a square pulse of duration $T$, for the parameters $\Delta =g^2  |\sin(k_F a)|$, $\delta =\beta =0$, and $g^2 T=1$.  The dotted (constant) line shows the corresponding probabilities for a single atom. The dashed line shows the result of including all the terms in the wavefunction, whereas the solid line shows only the contribution of the linear terms.}  %
\end{figure}

Now, it has been pointed out that the nonlinearity associated with the scattering of two photons from a V system could be used to build a conditional-phase (CPHASE) gate for quantum logic \cite{chudziki}, with only one substantial difficulty, namely, the entanglement generated in the course of the interaction.  Initial proposals to remove this entanglement and build up a large phase shift through successive interactions seemed rather unpractical \cite{chudziki}, but very recently Brod and Combes \cite{brod2,brod2} showed that disentanglement would happen spontaneously for counterpropagating photons in a maximally chiral arrangement (where each photon couples not only to just one transition, but also to only one direction of propagation).  The present system, therefore, might provide a perfect realization of the Brod-Combes proposal, since the photons naturally just pass through each other, with no reflection taking place. (Note, however, that the full phase gate would require concatenating several of these interaction nodes, each one formed by two two-level systems, along the waveguide, and a detailed study would need to be made of the response of such an arrangement to long, incoming, counterpropagating pulses.  We intend to pursue such a study in the very near future.)

The exact realization of the transformation (\ref{ne47}) in our system clearly requires relatively large, and rather precise, interactions between the atoms ($\Delta = g^2$).  It is, therefore, worth considering if an approximate alternative with $\Delta =0$ is possible.  The strong transmission features that we see in Figure 5b for $\delta = -g^2\tan\phi$ look like good candidates.  When this condition holds (and with $\Delta=0$), one has
\begin{align}
\Gamma_{\pm} = &\frac{g^2}{\cos\phi} e^{i\phi}\left( 1 \pm\cos\phi\right) \cr
\frac{\Gamma_c}{\Gamma_+} = &\frac{\Gamma_s}{\Gamma_-} = e^{-i\phi}\cos\phi
\end{align}
The last line implies that in the adiabatic (long-pulse) approximation (when $G_\pm \simeq f/\Gamma_\pm$), both the reflection coefficient $\rho(t)$ (Eq.~(\ref{ne41})) and the doubly-excited state contribution to $f_{ent,ab}^{+}$ (Eq.~(\ref{ne46})) will approximately vanish. In this limit (which requires that $\delta$ be of order $g^2$ or greater, so that both $\Gamma_+$ and $\Gamma_-$ will be sufficiently large), the scattered state (\ref{eq45}) becomes 
\begin{align}
&f_{aa}(t_1,t_2) = f_{bb}(t_1,t_2) \cr
&\;\simeq -\sqrt 2 \cos^2\phi\, e^{-2 i \phi} \left(e^{-\Gamma_+|t_1-t_2|} - e^{-\Gamma_-|t_1-t_2|} \right) f^2(t_<) \cr
%\sqrt{2} g^4}{\gamma^2} e^{-\gamma|t_1-t_2|}\sinh \Big[\frac{g^2\gamma |t_1-t_2|}{|\gamma|}\Big] f(t_<)^2  \cr
&f_{ab}(t_1,t_2) \simeq e^{-4i\phi} f(t_1)f(t_2) \cr
%\frac{\gamma*^2}{\gamma^2}f(t_1)f(t_2)\cr
&\;\;- 2 \cos^2\phi \,e^{-2 i \phi} \left(e^{-\Gamma_+|t_1-t_2|} + e^{-\Gamma_-|t_1-t_2|} \right) f^2(t_<)
\cr
\label{e48}
\end{align}
This shows that the reflected contribution can be made very small, so the two photons will still almost completely pass through each other (a result borne out by the numerical calculations shown in Fig.~8), whereas the transmitted term may still have a substantial nonlinear component, not very different from the one seen in Eq.~(\ref{ne47}).  Thus, in this limit, it would seem that the potential to make a CPHASE gate using the method of Brod and Combes also exists, even in the absence of a direct interaction between the atoms. We also intend to explore this possibility in detail very soon. 

\begin{figure}
\includegraphics[width=8cm]{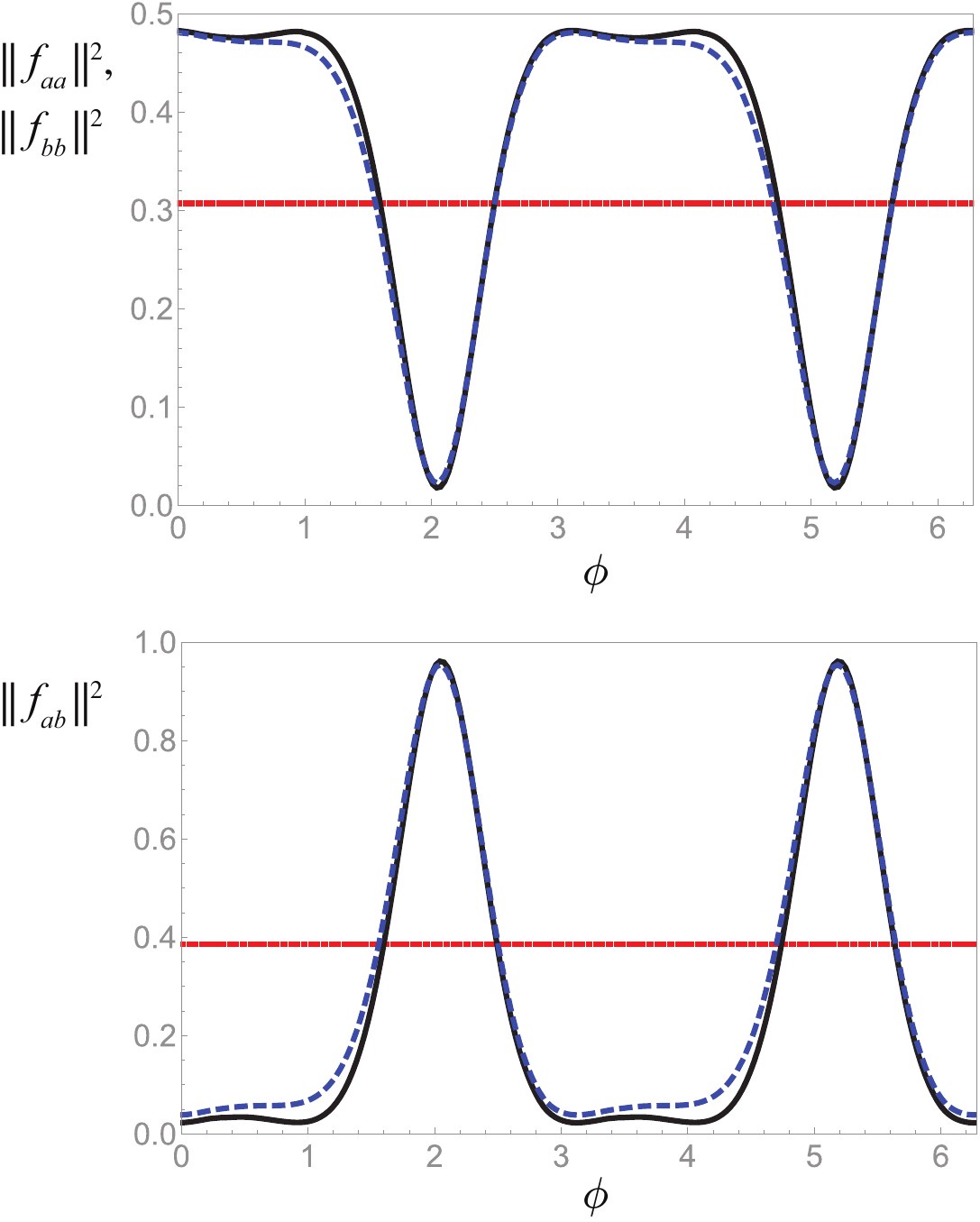}
\caption{\label{fig8} Probabilities of both photons leaving in the same direction (top) or in opposite directions (bottom), when the photons are initially counterpropagating. Values are calculated for a square pulse of duration $T$, for the parameters $\delta T=10$, $\Delta T=\beta T=0$, and $g^2 T=5$.  The dotted (constant) line shows the corresponding probabilities for a single atom. The dashed line shows the result of including all the terms in the wavefunction, whereas the solid line shows only the contribution of the linear terms.}  %
\end{figure}

\section{Conclusions}

In summary, we have studied analytically the photon transport and nonlinear interactions for a system of two two-level atoms maximally coupled to a lossless waveguide, for single- and two-photon pulses, allowing as well for direct interactions between the atoms (or two-level systems).  

We have shown that in the absence of interactions the linear, or single-photon, regime can be analyzed as an optical cavity.  This cavity exhibits a coupling and detuning that depends on the separation between the atoms and has well-separated resonances similar to a Fabry-Perot cavity. We have also shown that the atomic interactions, particularly the F\"orster-exchange-dipole term, can lead to very interesting behavior.  In particular, it allows the photon to perfectly transmit while acquiring  a frequency-dependent phase shift identical to a TLE in a unidirectional waveguide. 

We have also calculated the nonlinear effects due to the interaction between the two photons in the Markov approximation and identified situations where the two-atom and one-atom nonlinear contributions tend to cancel out,  allowing the system to effectively act as a linear scatterer.  Conversely, we have also identified a regime where the nonlinear contribution associated with the doubly-excited state would make it possible to use this system as a building block for a discriminator between one- and two-photon pulses. Finally, we have shown that it is possible for two counterpropagating photons in this system to interact with each other and remain in their respective modes. This last possibility warrants further study as it could be used as a nonlinearity in a passive CPHASE gate. 

\appendix*

\section{Details of some of the calculations}

Starting with the equations of motion, we redefine operators $C$, $D$, $\hat C^\prime$, $D^\prime$, so the equations now look like
\begin{subequations}
\begin{align}
\frac{d}{dt}\ket{\psi_{g}} &= -ig \left(\hat C^\dagger(t) \ket{\psi_{+}} + \hat D^\dagger(t) \ket{\psi_{-}} \right) \label{a1a} \\
\frac{d}{dt}\ket{\psi_{+}} &= -ig \left( \hat C(t) \ket{\psi_{g}} +  \hat {C^\prime}^\dagger(t) \ket{\psi_{e}} \right) \label{a1b} \\
\frac{d}{dt}\ket{\psi_{-}} &= -ig \left( \hat D(t) \ket{\psi_{g}} -  \hat {D^\prime}^\dagger(t) \ket{\psi_{e}} \right) \label{a1c} \\
\frac{d}{dt}\ket{\psi_{e}} &= -ig \left(\hat C^\prime(t) \ket{\psi_{+}} - \hat D^\prime(t) \ket{\psi_{-}} \right) \label{a1d}
\end{align} 
\label{a1}
\end{subequations}
To the extent the Markov approximation applies, the commutation relations for the new operators are the same as for the old ones. We shall not need commutators like  $[\hat C(t_1),\hat {C^\prime}^\dagger(t_2)]$ and $[\hat D(t_1),\hat {D^\prime}^\dagger(t_2)]$.  We will also use subscripts to indicate the time arguments, as in $\hat C_1 = \hat C(t_1)$. All integrals are assumed to be nested, with the largest subscript corresponding to the innermost time. The initial state is $\ket{\psi_0}$.

Formally integrating Eq.~(\ref{a1a}) we get
\begin{equation}
\ket{\psi_g} = \ket{\psi_0} -ig\int\left(\hat C^\dagger_1\ket{\psi_+} + \hat D^\dagger_1\ket{\psi_-}\right)
\label{a2}
\end{equation}
By substituting this in (\ref{a1b}), putting that equation in normal order, and performing the Markov approximation, as described in the main text we have
\begin{align}
\frac{d}{dt}\ket{\psi_{+}} 
&\simeq -\gamma_+ \ket{\psi_+} -ig \hat C \ket{\psi_0} -ig \hat {C^\prime}^\dagger\ket{\psi_{e}} \cr
&\qquad - g^2\int\left(\hat C^\dagger_1 C \ket{\psi_+} + \hat C\hat D^\dagger_1\ket{\psi_-}\right) 
\label{a3}
\end{align}
so
\begin{align}
\ket{\psi_{+}} = &-ig\int e^{-\gamma_+(t-t_1)} \left(\hat C_1 \ket{\psi_0} + \hat {C^\prime_1}^\dagger\ket{\psi_{e}}\right) \cr
& - g^2\int e^{-\gamma_+(t-t_1)}\int\left(\hat C^\dagger_2 \hat C_1 \ket{\psi_+} + \hat C_1\hat D^\dagger_2\ket{\psi_-}\right) \cr
\label{a4}
\end{align}
Similarly,
\begin{align}
\ket{\psi_{-}} = &-ig\int e^{-\gamma_-(t-t_1)} \left(\hat D_1 \ket{\psi_0} - \hat {D^\prime_1}^\dagger\ket{\psi_{e}}\right) \cr
& - g^2\int e^{-\gamma_-(t-t_1)}\int\left(\hat D^\dagger_2 \hat D_1 \ket{\psi_-} + \hat D_1\hat C^\dagger_2\ket{\psi_+}\right) \cr
\label{a5}
\end{align}
To solve the system (\ref{a4}), (\ref{a5}), we substitute both equations in their own right-hand sides.  After normal ordering, one finds that only the term containing $\ket{\psi_0}$ actually has a non-vanishing contribution within the limits of the Markov approximation. 

As an example of how to apply the Markov approximation, consider a term like
\begin{equation}
\int^t\int^{t_1}\int^{t_2}\int^{t_3} e^{-\gamma_+(t-t_1+t_2-t_3)} \hat C^\dagger_2 \hat C_1 \hat C^\dagger_4 \hat C_3 \ket{\psi_+}
\end{equation}
Normal-ordering the operators yields three delta function terms, $\delta(t_1-t_4)$, $\delta(t_1-t_4-a/c)$, and  $\delta(t_1-t_4+a/c)$. The first of these is nested too deep and will only be nonzero at one point. This arises from the fact that the integrals constrain the variables to be $t_1 \geq t_2 \geq t_3 \geq t_4$, but the delta function causes $t_4=t_1$. The $t_3$ integral is then only nonzero at one point, when $t_3=t_2=t_1$ at the upper limit, and thus the integral over $t_2$ is zero. The last term is also trivial as the delta function is only satisfied for $t_4=t_1+a/c$ but the integral ordering again constrains $t_4\leq t_1$ and as such the delta function is always zero. 

The middle term is nonzero but contributes little towards the overall solution and thus can be safely ignored. This delta function constrains the time variables so that $t_1\geq t_2\geq t_3\geq t_1-a/c$. In terms of $t_1$, then, the  limits of integration of the $t_3$ integral are $(t_1-a/c,t_1)$. Within the Markov approximation, that $t_1-a/c\approx t_1$, the limits of the integral are the same and as such the integral is zero. 

The consequence of this is that for any operators with time indices $t_i$ and $t_j$, when $|j-i|>1$ the operators effectively commute. With this approximation, and making use of the fact that $|\psi_{\pm}\rangle$ have only one photon each, whereas $|\psi_{e}\rangle$ has none, when we substitute (\ref{a4}), (\ref{a5}) into themselves and each other, only the first integral contributes.  One then arrives at
\begin{align}
\ket{\psi_{+}} = &-ig\int e^{-\gamma_+(t-t_1)} \left(\hat C_1 \ket{\psi_0} + \hat {C^\prime_1}^\dagger\ket{\psi_{e}}\right) \cr
& +i g^3\int\int \int e^{-\gamma_+(t-t_1)} e^{-\gamma_+(t_2-t_3)} \hat C^\dagger_2 \hat C_1 \hat C_3 \ket{\psi_0} \cr
& +i g^3\int\int \int e^{-\gamma_+(t-t_1)} e^{-\gamma_-(t_2-t_3)}\hat C_1\hat D^\dagger_2 \hat D_3 \ket{\psi_0}\cr
\label{a7}
\end{align}
\begin{align}
\ket{\psi_{-}} = &-ig\int e^{-\gamma_-(t-t_1)} \left(\hat D_1 \ket{\psi_0} - \hat {D^\prime_1}^\dagger\ket{\psi_{e}}\right) \cr
& +i g^3\int\int \int e^{-\gamma_-(t-t_1)} e^{-\gamma_-(t_2-t_3)} \hat D^\dagger_2 \hat D_1 \hat D_3 \ket{\psi_0} \cr
& +i g^3\int\int \int e^{-\gamma_-(t-t_1)} e^{-\gamma_+(t_2-t_3)}\hat D_1\hat C^\dagger_2 \hat C_3 \ket{\psi_0}\cr
\label{a8}
\end{align}
These equations are now to be substituted into Eq.~(\ref{a1d}), where, after putting the operators in normal order, it becomes apparent that none of the triple-integral terms contribute.  The $\ket{\psi_e}$ terms contribute a term $-(\gamma_+ +\gamma_-)\ket{\psi_e} = -2g^2\ket{\psi_e}$, which can be readily integrated, with the result
\begin{align}
\ket{\psi_e} &= -g^2 \int\int e^{-2g^2(t-t_1)} \cr
&\times\left(e^{-\gamma_+(t_1-t_2)}\hat C^\prime_1 \hat C_2 - e^{-\gamma_-(t_1-t_2)} \hat D^\prime_1 \hat D_2 \right)\ket{\psi_0}\cr
\label{a9}
\end{align}
At this point, all that is left is to substitute the result (\ref{a9}) into (\ref{a7}) and (\ref{a8}), and then substitute both of these into (\ref{a2}).  The final result, after restoring the original $\hat C$ and $\hat D$ operators, is Eq.~(\ref{e29}) of the text.

It is also straightforward to see, from Eq.~(\ref{a9}), that if we write $\ket{\psi_e} = \psi_e(t)\ket 0$ and if the field is initially in a ``$\hat C \hat C$'' state, the double excitation amplitude function $\psi_e(t)$ is equal to
\begin{equation}
\psi_e(t) = -e^{-i(2\delta + \beta)t} \sqrt 2\, g^2\cos^2(k_F a/2) \mathcal{E}_+(t)
\label{a10}
\end{equation}
where $\mathcal{E}_\pm(t)$ is the function defined in Eq.~(\ref{e34}) of the main text. We get a similar result (changing the cosine to a sine, and $\mathcal{E}_+$ to $\mathcal{E}_-$) if the initial state is of the ``$DD$'' type.

 \end{document}